# TAPPS Release 1: Plugin-Extensible Platform for Technical Analysis and Applied Statistics

**Justin Sam Chew**

**Colossus Technologies LLP, Republic of Singapore**
*chewjustinsam@gmail.com*

**Maurice HT Ling**

**Colossus Technologies LLP, Republic of Singapore**
**School of BioSciences, The University of Melbourne**
**Parkville, Victoria 3010, Australia**
*mauriceling@colossus-tech.com*

**Abstract**

We present the first release of TAPPS (Technical Analysis and Applied Statistics System); a Python implementation of a thin software platform aimed towards technical analyses and applied statistics. The core of TAPPS is a container for 2-dimensional data frame objects and a TAPPS command language. TAPPS language is not meant to be a programming language for script and plugin development but for the operational purposes. In this aspect, TAPPS language takes on the flavor of SQL rather than R, resulting in a shallower learning curve. All analytical functions are implemented as plugins. This results in a defined plugin system, which enables rapid development and incorporation of analysis functions. TAPPS Release 1 is released under GNU General Public License 3 for academic and non-commercial use. TAPPS code repository can be found at http://github.com/mauriceling/tapps.

## 1. Introduction

There are a number of statistical platforms available in the market today, which can be categorized in a number of different ways. In terms of licenses, some platforms are proprietary licensed (such as SAS [1] and Minitab [2]) while others are open source licensed (such as SOCR [3] and ELKI [4]). Between command-line interface (CLI) and graphical user interface (GUI), some are GUI-based (such as SOFA [5]) or CLI-based (such as ASReml [6]) while others (such as ELKI [4] and JMP [7]) have both GUI and CLI. Among the CLI-based platforms, some platforms (such as SciPy [8]) exist as libraries to known programming languages or develop their own Turing complete programming languages (such as R [9]) while others (such as TSP [10]) implements a domain-specific command language that does not aim to be a Turing complete language. The main advantage of domain-specific command language; of which, a popular example is SQL; is a shallower learning curve compared to a full programming language.

Here, we present TAPPS release 1 (abbreviation for "Technical Analysis and Applied Statistics System"; hereafter, known as "TAPPS") as a platform aimed towards technical analyses and applied statistics. TAPPS is licensed under GNU General Public License version 3 for academic and non-profit use. The main features of TAPPS are (1) a thin platform with (2) a CLI-based, domain-specific command language where (3) all analytical functions are implemented as plugins. This results in a defined plugin system, which enables rapid prototyping and testing of analysis functions. TAPPS code repository can be found at http://github.com/mauriceling/tapps.

## 2. Architecture and Implementation

Python programming language is chosen for TAPPS implementation due to its inherent language features [11], which promotes code maintainability. As of writing, one of the authors is also the project architect and main developer for COPADS (https://github.com/mauriceling/copads), a library of algorithms and data structures, developed entirely in Python programming language and has no third-party dependencies.

This section documents that architecture and implementation of TAPPS in a level of detail sufficient for interested developers to fork the code for further improvements.

TAPPS consists of 4 major sub-systems (Figure 1), supported by a third-party library. Users interact with

TAPPS though its command-line interface. The instructions are parsed into bytecode instructions using Python-Lex-Yacc [12]. The bytecodes are executed by TAPPS virtual machine to move data from the multi data frame object (data holder) to one or more plugins (functionalities), and load the results from plugins back to the multi data frame.

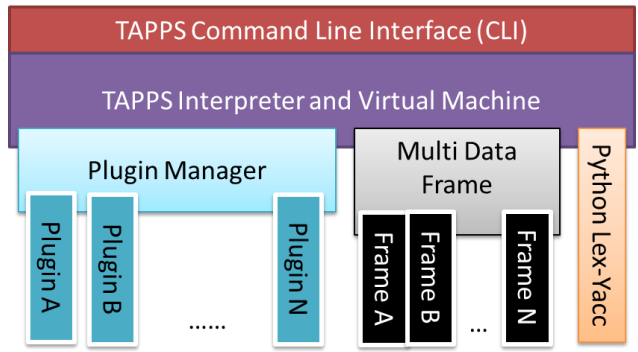

Fig. 1 Architecture of TAPPS.

**Multi Data Frame.** TAPPS uses the data frame object from COPADS (file: dataframe.py), which is essentially a 2-dimensional data table, as its main data structure (Figure 1). In practice, it is common to have more than one data frame object within a session of use. For example, the user can load one or more CSV files into TAPPS where each CSV file is maintained as a data frame. Furthermore, the user can slice a data frame into multiple data frames based on values or duplicate data frames. Hence, in order to accommodate for multiple data frames, COPADS' multi data frame object is used in TAPPS instead of using COPADS' data frame object directly as multi data frame object is a container for one or more data frame objects.

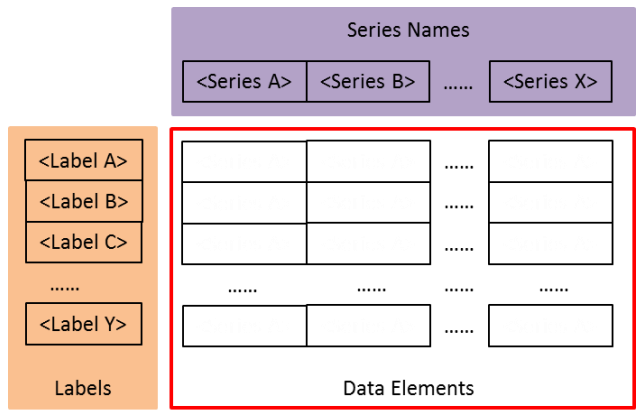

Fig. 2 Schematics of a Data Frame.

The core of a data frame (Figure 2) is a Python dictionary where the each data row is a value list with a unique key. This implies that each data row has a row label. Each data row within the same data frame must contain the same number of data elements. Each data element or value in a row is attributed to a series name, which implies that each series is a column in table form. This ensures that each data frame is either rectangular or square.

**Plugins and Plugin Manager.** As a thin platform, TAPPS has no analytical functions and does nothing in itself. All analytical functions, including statistical testing routines, are supplied via plugins. Hence, the operational richness of TAPPS is solely based on the repertoire of plugins. Each plugin is implemented as a standalone Python module, in the form of a folder consisting of at least 3 files – (1) the initialization file (file: __init__.py), which allows Python virtual machine to recognize the folder as an importable module; (2) a manifest file (file: manifest.py), which contains attributes to describe the plugin; and (3) an operation file (file: main.py), which must contain a "main" function as the execution entry point to the plugin. In addition, the "main" function takes a single "parameter" dictionary as input parameter, and returns a modified "parameter" dictionary back to TAPPS. All plugins must reside within "plugins" folder in TAPPS (please see Appendix B for a brief description on writing a plugin). At startup, the Plugin Manager will scan all individual sub-folders within the "plugins" folder as potential plugins, and attempts to load each potential plugin.

The "parameter" dictionary forms the communication link between TAPPS and its plugins; hence, an empty "parameter" dictionary for the specific plugin must be included in the main.py file. This is analogous to a customer (TAPPS) sending his/her laptop with instructions (the "parameter" dictionary) to a computer hardware shop (the plugin). The service staff in the hardware shop ("main" function of the plugin) pulls open the laptop, performs requested hardware and software upgrades or services (based on options in the "parameter" dictionary), then closes up the laptop and return the laptop with the service receipt/report as proof of initial service instruction to the customer (TAPPS). Hence, the "parameter" dictionary consists of three major components – the input data frame (the pre-serviced laptop), the results data frame (serviced laptop and service receipt/report), and a set of options that can vary with each plugin (the service requirements).

**Command Line Interface.** TAPPS command line interface (CLI) facilitates interaction between TAPPS and the user. The CLI can be launched in 2 different modes – interactive mode or script mode. In either mode, execution of the preceding command must complete before the next command can be executed (please see Appendix C for TAPPS language definition). In interactive mode, TAPPS behaves similarly to that of SAS [1] and R [9] CLI where the user enters each command sequentially, and executes in the order of entry.

In script mode, also commonly known as batch mode, a plain-text script written in TAPPS language to represent a sequential batch of commands is given to TAPPS. This enables plugins implementing computationally intensive statistical algorithms [13], such as Monte Carlo methods [14], to run a job without the need of user intervention. An important feature of a script is that a script can incorporate one or more scripts (Figure 3); thereby, allowing re-use and improving the maintenance of scripts. This is achieved through an "include" pre-processor, "@include", which is syntactically identical to that ("#include") in C language. Operationally, the "include" pre-processor represents the substitution location of another script file. Hence, it is possible for recursive inclusion of script files provided no cyclic inclusion is observed.

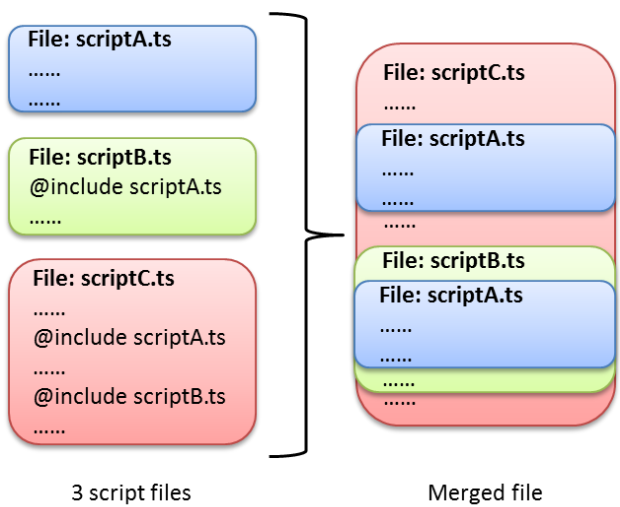

Fig. 3 Import and Merger of Script Files.

**Interpreter and Virtual Machine.** All TAPPS commands fed through the CLI, either as interactive mode or script mode, are processed by TAPPS interpreter (comprising of TAPPS lexer and parser, which uses Python-Lex-Yacc [12]) into TAPPS bytecodes (please see Appendix D for TAPPS bytecode definition). TAPPS bytecodes follow the *de facto* format [15] by beginning with an instruction (commonly known as opcode) and followed by zero or more operand(s).

The generated bytecodes are input to TAPPS virtual machine (TVM), which consists of 3 main components – (1) a set of environment variables implemented as a Python dictionary, which holds all the environmental settings; (2) a set of session variables implemented as a Python dictionary, which holds all data objects; and (3) TAPPS bytecode executor.

The set of session variables, which is implemented as a Python dictionary, consists of 3 main variables. Firstly, the multi data frame object (session['MDF']) to hold all data frames, which can be found within session['MDF'].frames. For example, a data frame with the ID of "DataA" will be accessible as session['MDF'].frames['DataA']. Secondly, all plugin parameter dictionaries are held as "parameters" (session['parameters']), which is implemented as a Python dictionary with the ID of the parameters as dictionary key. For example, a parameter set/dictionary with the ID of "pluginA_test" will be accessible as session['parameters']['pluginA_test']. Finally, all potential plugins will be recorded in session['plugins'], which is implemented as a Python dictionary. The list of successfully loaded plugins will be listed in session['plugins']['loaded'] while the list of plugins that fail to be loaded will be listed in session['plugins']['loadFail'].

For each successfully loaded plugin, a new dictionary, with "plugin_<plugin name>" as key, will be created in session to contain information and the parameter dictionary associated with the plugin. For example, if a "ztest" plugin is successfully loaded, information regarding "ztest" plugin will be accessible as session['plugin_ztest']. At the same time, session['plugin_ztest']['main'] will point to ztest's main function, and ztest's parameters dictionary can be found at and be copied from session['plugin_ztest']['parameters'].

TAPPS bytecode executor is implemented as large IF statement, which channels the bytecode operands (if any) to their respective supportive executors based on the opcode. Each supportive executor then takes the given operands and operates on the session variables and environment variables.

## 3. TAPPS Language

TAPPS Language defines 13 types of statements (Appendix C) and can be separated into 2 sections – meta-commands and operatives. Meta-command statements are mainly variables setting and display commands, used to manage the behavior or provide information with regards to the current session. The following explains the 4 meta-command statements:

- Describe: describe a data object or variable, which is more informative than "show" statement

- Set: set options in environment or within plugin parameters dictionary
- Shell: launch a language sub-shell or execute a non-TAPPS statement
- Show: show values of the session or environment variable/item.

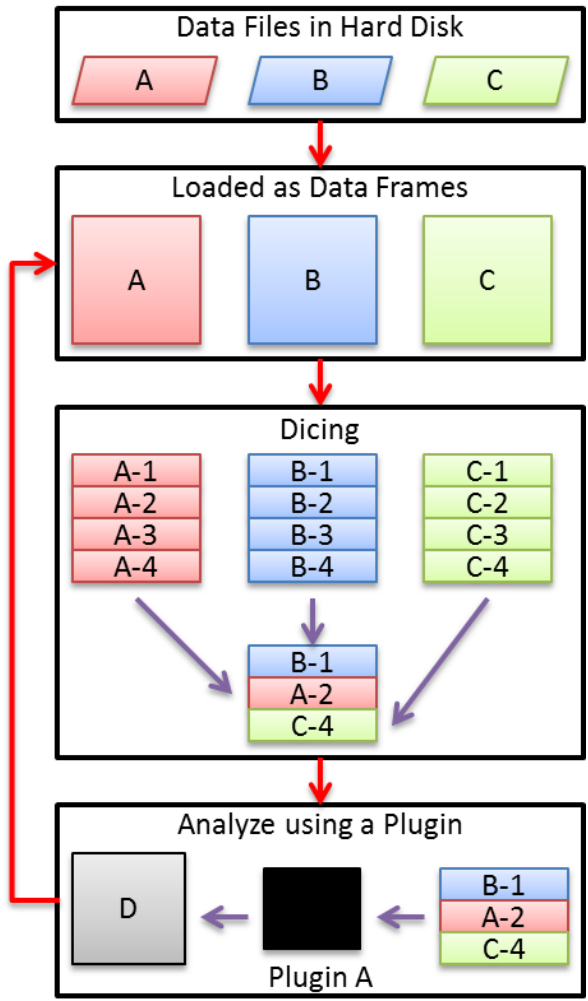

Fig. 4 Load-Dice-Analyze Cycle. Three separate data files (A, B, and C) are loaded into separate data frames (A, B, and C). Each data frame is divided into 4 smaller data frames; for example, Data Frame A is divided into Data Frames A-1, A-2, A-3, and A-4. One data frame from each of the original file (B-1, A-2, and C-4) are merged into a data frame, which is then analyzed by Plugin A. The output of Plugin A is Data Frame D, which can be attached as a data frame for the next cycle.

The operative statements manage the functionalities and operations of TAPPS; more specifically, the Load-Dice-Analyze cycle (Figure 4). Under this cycle, data is first loaded into TAPPS by reading in one or more data files. The next step is Dice, which is the extraction part of the loaded data and/or merge data segments together. For example, several loaded CSV files into data frames can be sliced into multiple smaller data frames by selection criteria before logically merging the sliced data frames into a larger data frame. Finally, the merged data frame can be analyzed by perform technical and/or statistical analyses using a plugin. As the output result of an analysis by plugin is a data frame object, it can be attached into the multi data frame object for a repeat of the Load-Dice-Analyze cycle.

The following explains the 9 operative statements:

- Cast: type cast values into a different data type
- Delete: delete a data frame from multi data frame object, or a plugin parameter set
- Load: load an external file or saved session
- Merge: merge 2 data frames by series or labels
- New: duplicate a plugin parameter set or attach a data frame within a plugin parameter set as a new data frame object
- Rename: rename series or label names within a data frame
- Runplugin: performs an analysis by running a plugin
- Save: save a session
- Select: generate a new data frame from a selection criterion on an existing data frame

The following is an example of a session:

1. Set up the environment (set current working directory to "data" directory relative to the startup directory)

```
TAPPS: 1> set rcwd data
```

2. Load a CSV file (load <TAPPS directory>/data/STI_2015.csv and attach the data frame as "STI")

```
TAPPS: 2> load csv STI_2015.csv as STI
```

3. Dice the data (type cast "Open" to from string to float; slice 2 data frames, STI_Low and STI_High, from STI; and merge STI_High and STI_Low into STI_A)

```
TAPPS: 3> cast Open in STI as nonalpha
TAPPS: 4> select from STI as STI_Low where
    Open < 820
TAPPS: 5> select from STI as STI_High where
    Open > 2000
TAPPS: 6> select from STI_Low as STI_A
TAPPS: 7> merge labels from STI_High to
    STI_A
```

4. Set up a plugin parameter dictionary (prepare plugin parameter dictionary from "summarize" plugin, name the parameter dictionary as "testingA", and set to run "by_series" method on "STI_A" data frame)

```
TAPPS: 8> new summarize parameter as
    testingA
TAPPS: 9> set parameter analysis_name in
    testingA as trialA
TAPPS: 10> set parameter analytical_method
    in testingA as by_series
TAPPS: 11> set parameter dataframe in
    testingA as STI_A
```

5. Run the plugin

```
TAPPS: 12> runplugin testingA
```

6. Attach results data frame

```
TAPPS: 13> new STI_summarize dataframe from
    testingA results
TAPPS: 14> show dataframe

Current Dataframe(s) (n = 5):

  Dataframe Name: STI_High
  Series Names:
Open,High,Low,Close,Volume,Adj Close
  Number of Series: 6
  Number of Labels (data rows): 3919

  Dataframe Name: STI_Low
  Series Names:
Open,High,Low,Close,Volume,Adj Close
  Number of Series: 6
  Number of Labels (data rows): 3

  Dataframe Name: STI_A
  Series Names:
Open,High,Low,Close,Volume,Adj Close
  Number of Series: 6
  Number of Labels (data rows): 3922

  Dataframe Name: STI
  Series Names:
Open,High,Low,Close,Volume,Adj Close
  Number of Series: 6
  Number of Labels (data rows): 6996

  Dataframe Name: STI_summarize
  Series Names:
arithmetic_mean,count,maximum,median,minimu
m,standard_deviation,summation
  Number of Series: 7
  Number of Labels (data rows): 6

TAPPS: 15> describe STI_summarize

Describing Dataframe - STI_summarize

  Series Names:
arithmetic_mean,count,maximum,median,minimu
m,standard_deviation,summation
  Number of Series: 7
  Number of Labels (data rows): 6

  Series Name - arithmetic_mean
  Minimum value in arithmetic_mean:
2673.26085131
  Maximum value in arithmetic_mean:
92459705.7114
  Number of string data type values: 0
  Number of integer data type values: 0
  Number of float data type values: 6
  Number of unknown data type values: 0

------- <truncated> ------
```

7. Change current working directory, save a data frame and session before exit (change current working directory to serialize session as <TAPPS directory>/examples, saves STI_A data frame as a CSV file, and serialize current session to <TAPPS directory>/examples /tapps_manuscript.txt)

```
TAPPS: 16> set ocwd
TAPPS: 17> set rcwd examples
TAPPS: 18> save dataframe STI_A as csv
    STI_A.csv
TAPPS: 19> save session as
    tapps_manuscript.txt
TAPPS: 20> exit
```

## 4. Conclusion and Future Work

In this article, we present TAPPS as a plugin extensible tool for technical analysis and applied statistics, which is driven by a domain-specific command-line language. This article documents the architecture and implementation of TAPPS to document essential concepts, which are needed for further development. An example session of TAPPS use has been given as illustration.

The immediate future work is to develop plugins for TAPPS. For example, hypothesis test routines what are implemented in COPADS [16, 17] and existing Python implementations of econometrical methods [18] can be wrapped into plugins.

### Acknowledgement
The authors wish to thank T. Yeo (Union Bank of Switzerland), J. Oon (University of Queensland), and C. Kuo (Yahoo) for their comments on the initial drafts of this manuscript.

## Appendix A: List of Statistical Package Available for All 5 Major Operating Systems

A list of 15 statistical packages, out of 57 listed statistical packages (https://en.wikipedia.org/wiki/Comparison_of_statistical_packages; accessed), are available for all 5 major operating systems; namely, Microsoft Windows, Mac OSX, Linux, BSD, and UNIX. Of these 15 packages that are available for all 5 major operating systems, 12 are open source packages. Four of the 12 open source packages (DataMelt, Dataplot, SciPy and Statsmodels) do not have a defined plugin system but by their exposure of internal APIs, presented an implicit plugin system. Of these 12 open source packages, only 2 packages (R and ROOT) have defined plugin system for addition of statistical analysis packages into the system.

| Statistical Package | Open Source | License | Interface | Scripting | Plugin System |
|---|---|---|---|---|---|
| R | Yes | GNU GPL | CLI/GUI | R, Python, Perl | Yes |
| ROOT | Yes | GNU GPL | GUI | C++, Python | Yes |
| DataMelt | Yes | GNU GPL | CLI/GUI | Java, Jython, Groovy, Jruby, BeanShell | Partial |
| Dataplot | Yes | Public Domain | CLI/GUI | None | Partial |
| SciPy | Yes | BSD | CLI | Python | Partial |
| Statsmodels | Yes | BSD | CLI | Python | Partial |
| ADaMSoft | Yes | GNU GPL | CLI/GUI | None | No |
| OpenEpi | Yes | GNU GPL | GUI | None | No |
| PSPP | Yes | GNU GPL | CLI/GUI | Perl | No |
| Salstat | Yes | GNU GPL | CLI/GUI | Python | No |
| SOCR | Yes | GNU LGPL | GUI | None | No |
| SOFA Statistics | Yes | AGPL | GUI | Python | No |
| Statwing | No | Proprietary | GUI | None | Yes |
| Brightstat | No | Proprietary | GUI | None | No |
| TSP | No | Proprietary | CLI | None | No |

## Appendix B: How to Write a Plugin

A TAPPS plugin is a Python module with the following rules:

1. A plugin has to be a folder named in the following format: `<plugin name>_<release number>`, where `<plugin name>` has to be a single word in small caps and without underscore or hyphenation, and `<release number>` has to be an integer. For example, "generallinearmodel_1" is allowed while "General_Linear_Model_1" or "glm_1.0" or "GeneralLinearModel_1" are not allowed.
2. There must be the following files within the plugin folder – (1) `__init__.py`, (2) `manifest.py`, and (3) `main.py`.
3. Items in `manifest.py` must be filled in, as required.
4. The main file, `main.py`, must contain 3 items – (1) `instruction` string (a multi-line string to explain to users how to fill in the parameters dictionary); (2) a dictionary called `parameters`; and (3) a function called `main`.
5. The `parameters` dictionary must minimally contain the following key-value pairs – (1) `plugin_name`, which must correspond to the name of the plugin; (2) `dataframe`, for attaching a data frame as input to the plugin; and (3) `results`, for the plugin to attach a data frame as analysis output. The following key-value pairs are not mandatory but encouraged – (1) `analysis_name`, for user to give a short description of the analysis; (2) `analytical_method`, for the plugin to know which analysis function to execute is there is more than one analytical option; and (3) `narrative`, for user to give a long description of the analysis.
6. The `main` function (1) must take only `parameters` dictionary as the only parameter, (2) use only information provided in `parameters` dictionary for execution in order to maintain encapsulation, (3) must return the analysis output as a data frame into `results` key of `parameters` dictionary, and (4) must return `parameters` dictionary at the end.
7. The `main` function can call other functions or other modules.

**A plugin template (https://github.com/mauriceling/tapps/tree/master/plugins/template_1), containing the essential boilerplate codes of a plugin has been provided. Plugin developers are encouraged to duplicate this folder as basic template for development.**

The following code is a section of main file (`main.py`) from 'summarize' plugin (https://github.com/mauriceling/tapps/tree/master/plugins/summarize_1):

```python
instructions = '''
```

```python
    How to fill in parameters dictionary (p):
    Standard instructions:
        p['analysis_name'] = <user given name of analysis in string>
        p['narrative'] = <user given description of analysis, if any>
        p['dataframe'] = <input dataframe object from session['MDF'].frames>
    Instructions specific to this plugin:
        1. p['analytical_method'] takes in either 'by_series' (summarize by
        series) or 'by_labels' (summarize by labels).'''

parameters = {'plugin_name': 'summarize',
              'analysis_name': None,
              'analytical_method': None,
              'dataframe': None,
              'results': None,
              'narrative': None}

def main(parameters):
    '''Entry function for the 'summarize' plugin.

    @param parameters: set of parameters, including data frame, which are
    needed for the plugin to execute
    @type parameters: dictionary
    @return: parameters
    @rtype: dictionary'''
    # Step 1: Pull out needed items / data from parameters dictionary
    method = parameters['analytical_method']
    dataframe = parameters['dataframe']
    results = parameters['results']
    # END Step 1: Pull out needed items / data from parameters dictionary

    # Step 2: Perform plugin operations
    if method == 'by_series' or method == None:
        (statistics, labels) = summarize_series(dataframe)
        results.addData(statistics, labels)
    if method == 'by_labels':
        (data, series) = summarize_labels(dataframe)
        results.data = data
        results.label = data.keys()
        results.series_names = series
    # END Step 2: Perform plugin operations

    # Step 3: Load dataframe and results back into parameters dictionary
    parameters['dataframe'] = dataframe
    parameters['results'] = results
    # END Step 3: Load dataframe and results back into parameters dictionary
    return parameters

def summarize_series(dataframe):
    '''Function to statistical summaries of each series in the dataframe.'''
    labels = dataframe.data.keys()
    series = dataframe.series_names
    statistics = {'summation': [], 'arithmetic_mean': [],
                  'standard_deviation': [], 'maximum': [],
                  'minimum': [], 'median': [], 'count': []}
    for index in range(len(series)):
        sdata = [float(dataframe.data[label][index]) for label in labels]
        statistics['summation'].append(sum(sdata))              # 1. summation
        arithmetic_mean = float(sum(sdata)) / len(sdata)        # 2. arithmetic mean
        statistics['arithmetic_mean'].append(arithmetic_mean)
        index = int(len(sdata) / 2)                             # 3. median
        statistics['median'].append(sdata[index])
        sdata.sort()                                            # 4. maximum
```

```python
            statistics['maximum'].append(sdata[-1])
            statistics['minimum'].append(sdata[0])              # 5. minimum
            statistics['count'].append(len(sdata))              # 6. count
            if len(sdata) < 30:                                 # 7. standard deviation
                s = float(sum([(x-arithmetic_mean)**2 for x in sdata])) / (len(sdata) - 1)
            else:
                s = float(sum([(x-arithmetic_mean)**2 for x in sdata])) / len(sdata)
            statistics['standard_deviation'].append((s**0.5))
    return (statistics, series)

def summarize_labels(dataframe):
    '''Function to statistical summaries of each label in the dataframe.'''
    series = ['arithmetic_mean', 'count', 'maximum', 'median', 'minimum',
              'standard_deviation', 'summation']
    data = {}
    for label in dataframe.data.keys():
        sdata = [float(x) for x in dataframe.data[label]]
        temp = ['arithmetic_mean', 'count', 'maximum', 'median', 'minimum',
                'standard_deviation', 'summation']
        temp[0] = float(sum(sdata)) / len(sdata)     # 1. arithmetic mean
        temp[1] = len(sdata)                         # 2. count
        sdata.sort()                                 # 3. maximum
        temp[2] = sdata[-1]
        temp[3] = sdata[int(len(sdata) / 2)]         # 4. median
        temp[4] = sdata[0]                           # 5. minimum
        if (len(sdata) < 30) and (len(sdata) > 1):   # 6. standard deviation
            s = float(sum([(x-temp[0])**2 for x in sdata])) / (len(sdata) - 1)
        elif len(sdata) == 1:
            s = 'NA'
        else:
            s = float(sum([(x-temp[0])**2 for x in sdata])) / len(sdata)
        temp[5] = s ** 0.5
        temp[6] = sum(sdata)                         # 7: summation
        data[label] = [x for x in temp]
    return (data, series)
```

## Appendix C: TAPPS Language (Release 1) Definition

The following is TAPPS language (release 1) definition in Backus-Naur Form where reserved words are in bold:

```
statement : cast_statement | delete_statement | describe_statement | load_statement
          | merge_statement | new_statement | rename_statement | runplugin_statement
          | save_statement | select_statement | set_statement | shell_statement
          | show_statement
cast_statement : CAST id_list IN ID AS datatype
delete_statement : DELETE DATAFRAME ID | DELETE PARAMETER ID
describe_statement : DESCRIBE ID
load_statement : LOAD CSV FILENAME AS ID | LOAD NOHEADER CSV FILENAME AS ID
               | LOAD SESSION FROM FILENAME
merge_statement : MERGE SERIES id_list FROM ID TO ID
                | MERGE LABELS FROM ID TO ID | MERGE REPLACE LABELS FROM ID TO ID
new_statement : NEW ID PARAMETER AS ID | NEW ID DATAFRAME FROM ID plocation
rename_statement : RENAME SERIES IN ID FROM id_value TO id_value
                 | RENAME LABELS IN ID FROM id_value TO id_value
runplugin_statement : RUNPLUGIN ID
save_statement : SAVE SESSION AS FILENAME | SAVE DATAFRAME ID AS CSV FILENAME
select_statement : SELECT FROM ID AS ID | SELECT FROM ID AS ID WHERE binop value
                 | SELECT FROM ID AS ID WHERE ID binop value
set_statement : SET DISPLAYAST ID | SET CWD FOLDER | SET SEPARATOR separators
              | SET FILLIN fillin_options | SET PARAMETER ID IN ID AS ID
              | SET PARAMETER DATAFRAME IN ID AS ID | SET RCWD ID | SET OCWD
shell_statement : PYTHONSHELL
```

```
show_statement : SHOW ASTHISTORY | SHOW ENVIRONMENT | SHOW HISTORY | SHOW PLUGIN LIST
               | SHOW PLUGIN ID | SHOW SESSION | SHOW DATAFRAME | SHOW PARAMETER
id_list : ALL | ID | id_list DELIMITER ID
value : number_value | id_value
binop : DELIMITER | GE | LE | EQ | NE
datatype : ALPHA | NONALPHA | FLOAT | REAL | INTEGER
fillin_options : NUMBER | ID
id_value : ID | STRING
number_value : NUMBER
plocation : RESULTS | DATAFRAME
separators : DELIMITER | COMMA | COLON | SEMICOLON | RIGHTSLASH | BAR | DOT | PLUS | MINUS
           | TIMES | DIVIDE | GT | LT
```

## Appendix D: TAPPS Bytecode (Release 1) Description

The following table describes TAPPS bytecodes (release 1):

| Opcode | Operand(s) | Description |
|---|---|---|
| cast | <data type>, <series names>, <data frame name> | Type cast all data elements for a specific set of series names within a data frame. |
| deldataframe | <data frame name> | Deletes data frame from multi data frame object. |
| delparam | | Deletes plugin parameter(s) dictionary. |
| describe | <data frame name> | Describe a data frame – more informative than "show". |
| duplicateframe | <original data frame name>, <new data frame name> | Duplicate/deep-copy an existing data frame and attach the copy into multi data frame object with a new data frame name. |
| greedysearch | <data frame name>, <new data frame name>, <binary operator>, <value> | Select data from a data frame into another data frame, where a data element in any series, identified by the same data label, holds true for the selection criterion. |
| idsearch | <data frame name>, <new data frame name>, <series name>, <binary operator>, <value> | Select data from a data frame into another data frame, where a data element in a specific series, holds true for the selection criterion. |
| loadcsv1 | <file name>, <data frame name> | Load CSV file, with headers as series name, and attach the data frame. |
| loadcsv2 | <file name>, <data frame name> | Load CSV file, without headers, and attach the data frame. |
| loadsession | <file name> | Loads a previously saved session from file. |
| mergelabels1 | <source data frame name>, <destination data frame name> | Merge data, without replacing any data in destination data frame, from source data frame to destination data frame. |
| mergelabels2 | <source data frame name>, <destination data frame name> | Merge data from source data frame to destination data frame. Existing data within the destination data frame will be replaced with data from source data frame, if the label(s) in the destination data frame is/are found in the source data frame. |
| mergeseries | <series names>, <source data frame name>, <destination data frame name> | Merge data from one or more series from source data frame to destination data frame. |
| newdataframe | <new data frame name>, , {dataframe | results} | Duplicate/deep-copy the input data frame or output results from a parameters dictionary, and attach as a new data frame. |
| newparam | <plugin name>, <new parameter name> | Duplicate/deep-copy parameters dictionary, for use, from a loaded plugin. |
| pythonshell | No operand | Launch a Python interactive shell. |
| renameseries | <data frame name>, <original series names>, <new series names> | Rename series within a data frame. |
| renamelabels | <data frame name>, <original label names>, <new label names> | Rename label(s) within a data frame. |
| runplugin | | Run plugin using parameters dictionary; name of the plugin is coded within the parameters dictionary. |
| savecsv | <data frame name>, <file name> | Saves a data frame into a CSV file. |
| savesession | <file name> | Saves the current session into a file. |
| set | <variable>, <value> | Set options in environment or within plugin parameters dictionary. |
| show | <item> | Show values of the session or environment variable/item. |